\documentclass[times, twoside, watermark]{zHenriquesLab-StyleBioRxiv}

\leadauthor{Singh} 

\usepackage{amsmath}
\usepackage{bm}

\title{TopoSPAM: Topology grounded Simulation Platform for morphogenesis and biological Active Matter}
\shorttitle{TopoSPAM}

\author[1,2]{Abhinav Singh}
\author[1,2]{Abhijeet Krishna}
\author[3]{Aboutaleb Amiri}
\author[1,2,3]{Anne Materne}
\author[1,2]{Pietro Incardona}
\author[3,4]{Charlie Duclut}
\author[1,2]{Carlos M. Duque}
\author[1,2]{Alicja Sza\l{}apak}
\author[1,2]{Mohammadreza Bahadorian}
\author[1,2]{Sachin Krishnan Thekke Veettil} 
\author[1,2,5]{Philipp H. Suhrcke}
\author[2,3,6]{Frank J\"{u}licher}
\author[1,2,5,6,*]{Ivo F. Sbalzarini}
\author[1,2,6,*]{Carl D. Modes}

\affil[1]{Max Planck Institute of Molecular Cell Biology and Genetics, Pfotenhauerstr. 108, 01307 Dresden, Germany}
\affil[2]{Center for Systems Biology Dresden, Pfotenhauerstr. 108b, 01307 Dresden, Germany}
\affil[3]{Max Planck Institute for the Physics of Complex Systems, N\"{o}thnitzer Str. 38, 01187 Dresden, Germany}
\affil[4]{Laboratoire Physique des Cellules et Cancer (PCC), CNRS UMR 168, Institut Curie, Universit\'{e} Paris Sciences et Lettres, Sorbonne Universit\'{e}, Paris 75005, France}
\affil[5]{Dresden University of Technology, Faculty of Computer Science, N\"{o}thnitzer Str. 43, 01067 Dresden, Germany}
\affil[6]{Physics of Life Cluster of Excellence, TU Dresden, Arndoldstr. 18, 01307 Dresden, Germany}

\begin{document}

\maketitle

\begin{abstract}
    We present a topology grounded, multiscale simulation platform for morphogenesis and biological active matter. Morphogenesis and biological active matter represent keystone problems in biology with additional, far-reaching implications across the biomedical sciences. Addressing these problems will require flexible, cross-scale models of tissue shape, development, and dysfunction that can be tuned to understand, model, and predict relevant individual cases. Current approaches to simulating anatomical or cellular subsystems tend to rely on static, assumed shapes. Meanwhile, the potential for topology to provide natural dimensionality reduction and organization of shape and dynamical outcomes is not fully exploited. TopoSPAM combines ease of use with powerful simulation algorithms and methodological advances, including active nematic gels, topological-defect-driven shape dynamics, and an active 3D vertex model of tissues. It is capable of determining emergent flows and shapes across scales.  
\end{abstract}

Computational models of active, biological systems are rapidly growing in importance and fidelity in fundamental research but also in translation to the clinic and to in bio-inspired engineering. Such models can open up a new understanding of the mechanics and also the mechanistics of morphogenesis.

\begin{figure*}[h]
    \centering
    \includegraphics[width=17cm]{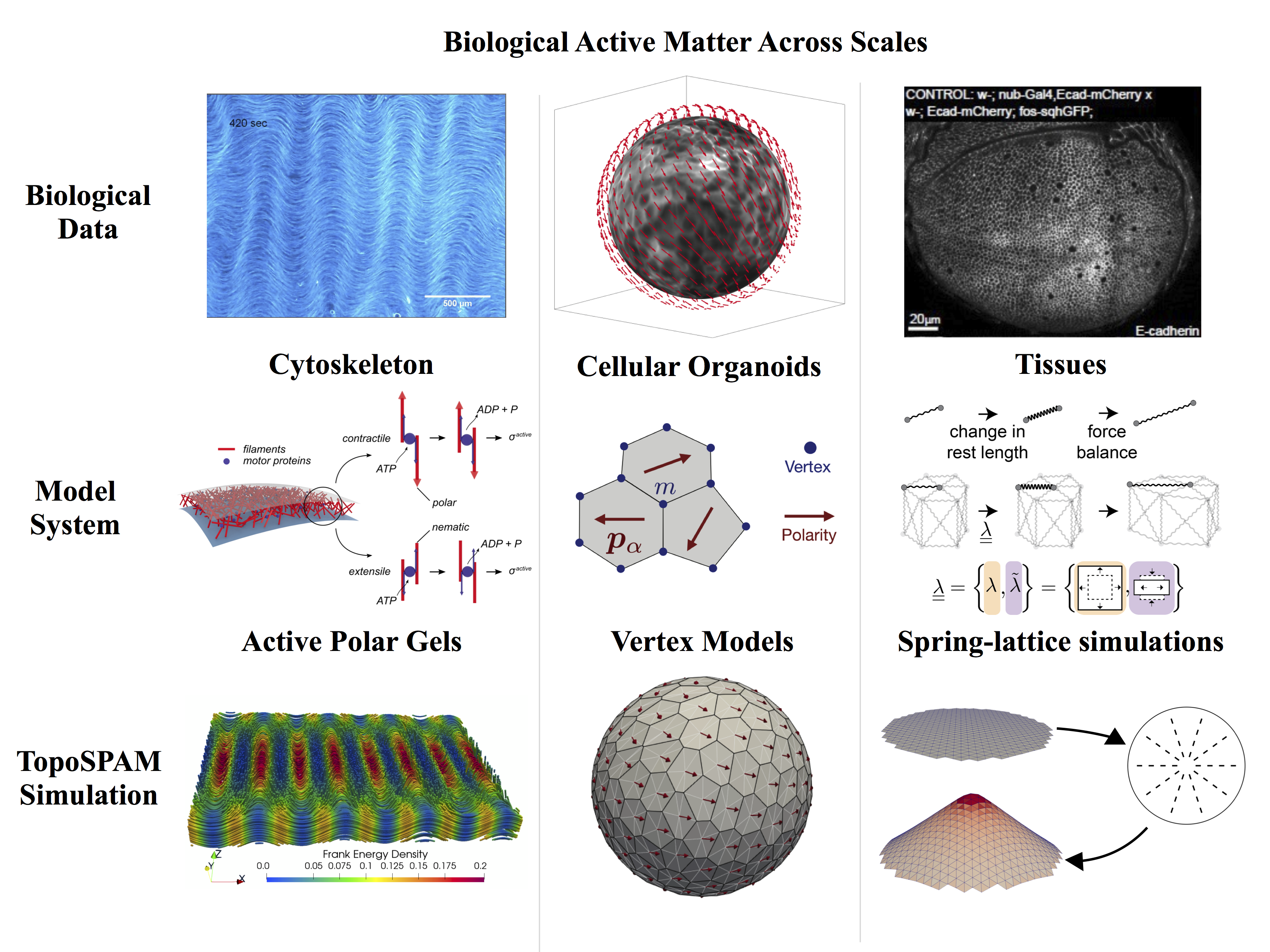}
    \caption{Modeling and simulation of biological active matter across scales. \textbf{Cytoskeleton (left column):} Microtubule mixtures with Kinesin-1 motor proteins exhibiting a bend instability (top panel)~\cite{PhysRevLett.125.257801}. An active polar fluid model (middle) captures the instability in the corresponding TopoSPAM Simulation (bottom panel).
    \textbf{Cellular organoids (center column):} Pancreatic organoids exhibiting spontaneous chiral rotations that are modeled with the 3D vertex model (top and middle)~\cite{Tan:2024}. The TopoSPAM simulation of the vertex model predicts the spontaneous chiral flows (bottom).
    \textbf{Tissue 3D shape change (right column):} {\it Drosophila} wing disc development (top)~\cite{10.7554/eLife.57964} modeled via a spring-lattice model (middle). The TopoSPAM simulation of the spring-lattice model predicts the three-dimensional development of the wing disc (bottom). Notice that the models are not restricted to the scales shown here. In particular, the active polar gel model can also be used across scales.}
    \label{fig:vertexmodel}
\end{figure*}

Robust approaches to molecular dynamics and Markov chain or density functional Monte Carlo have long presented researchers with a coherent set of tools for simulation on the molecular level. In contrast, morphogenesis is an inherently multiscale problem, spanning the intracellular, cellular, tissue, and organ mesoscales (Fig.~\ref{fig:vertexmodel}, top row). Studying biological morphogenesis is also a  multidisciplinary challenge, requiring combined advances in computer science, mathematics, physics, and biology (Fig.~\ref{fig:vertexmodel}, center row).

Here, we combine expertise from these disciplines for simulating active biological systems across the different mesoscales and introduce TopoSPAM: Topologically grounded Simulation Platform for morphogenesis and biological Active Matter (Fig.~\ref{fig:vertexmodel}, bottom row)~\cite{TopoSPAM}. Our approach is based on the observation that coarse-graining from scale to scale inevitably gives rise to emergent order parameters that support topologically non-trivial configurations, such as topological defects, grain boundaries, or domains. By combining powerful numerical methods and solvers with explicit consideration of these topological states, we can capture how they emerge, or use them as initial and boundary conditions to drive the simulations robustly in an effectively lower-dimensional phase space.

TopoSPAM leverages meshfree particle methods, where particles do not necessarily represent physical entities, such as molecules or cells, but rather mathematical collocation points on which continuous fields are discretized. This simplifies simulations in complex-shaped and moving or deforming geometries as are hallmarks of morphogenesis. It also supports both fixed and co-moving discretizations, allowing the user to control the trade-off between numerical stability and computational cost. TopoSPAM combines these continuous models with a quasi-3D, active, discrete vertex model, together with spring-lattice methods to capture elastic and visco-elastic shape programming. In these shape-programming methods, the simulated meshwork responds to spontaneous strain patterns represented by alterations to the meshwork structure. These changes are then capable of driving prescribed, non-trivial shape outcomes.

\section*{Results}

In TopoSPAM, we provide a unified framework for simulation at different biological scales by focusing on the classes of emergent order parameters and their attendant topological degrees of freedom likely to appear at the different scales. For cytoskeletal, or more general intracellular simulations, we provide a complete suite of tools for polar fluids, gels, and nematic liquid crystals -- in each case both active and passive versions. At the tissue scale, TopoSPAM addiitonally provides both a quasi-3D, active vertex model on curved surfaces, and coarse-grained 3D models capable of visco-elastic spontaneous-strain-driven shape programming simulations of morphogenesis.

\begin{figure*}[h]
    \centering
    {\bf (A)} \includegraphics[angle=-90,width=13.5cm]{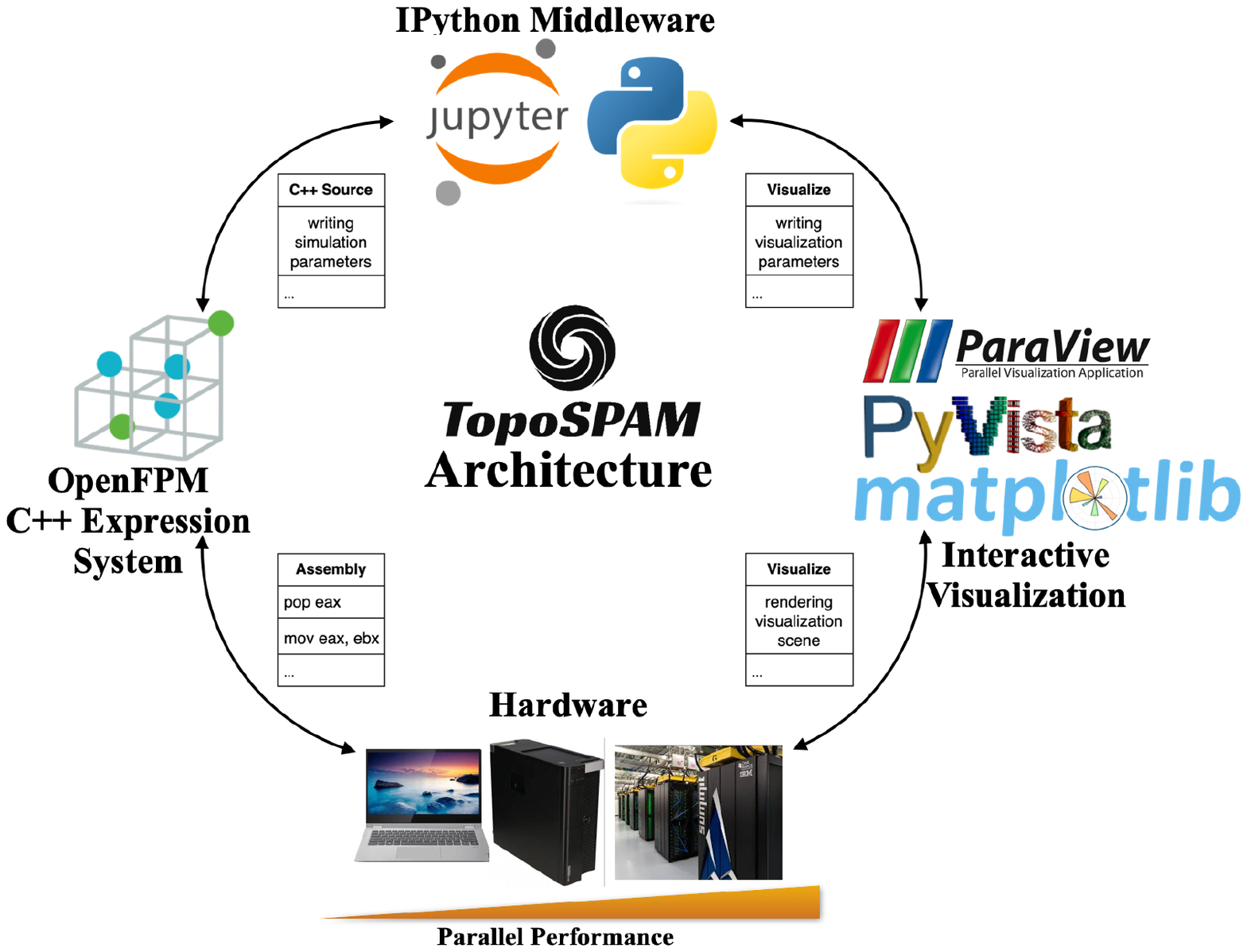}\\
    {\bf (B)} \includegraphics[angle=-90,width=13.5cm]{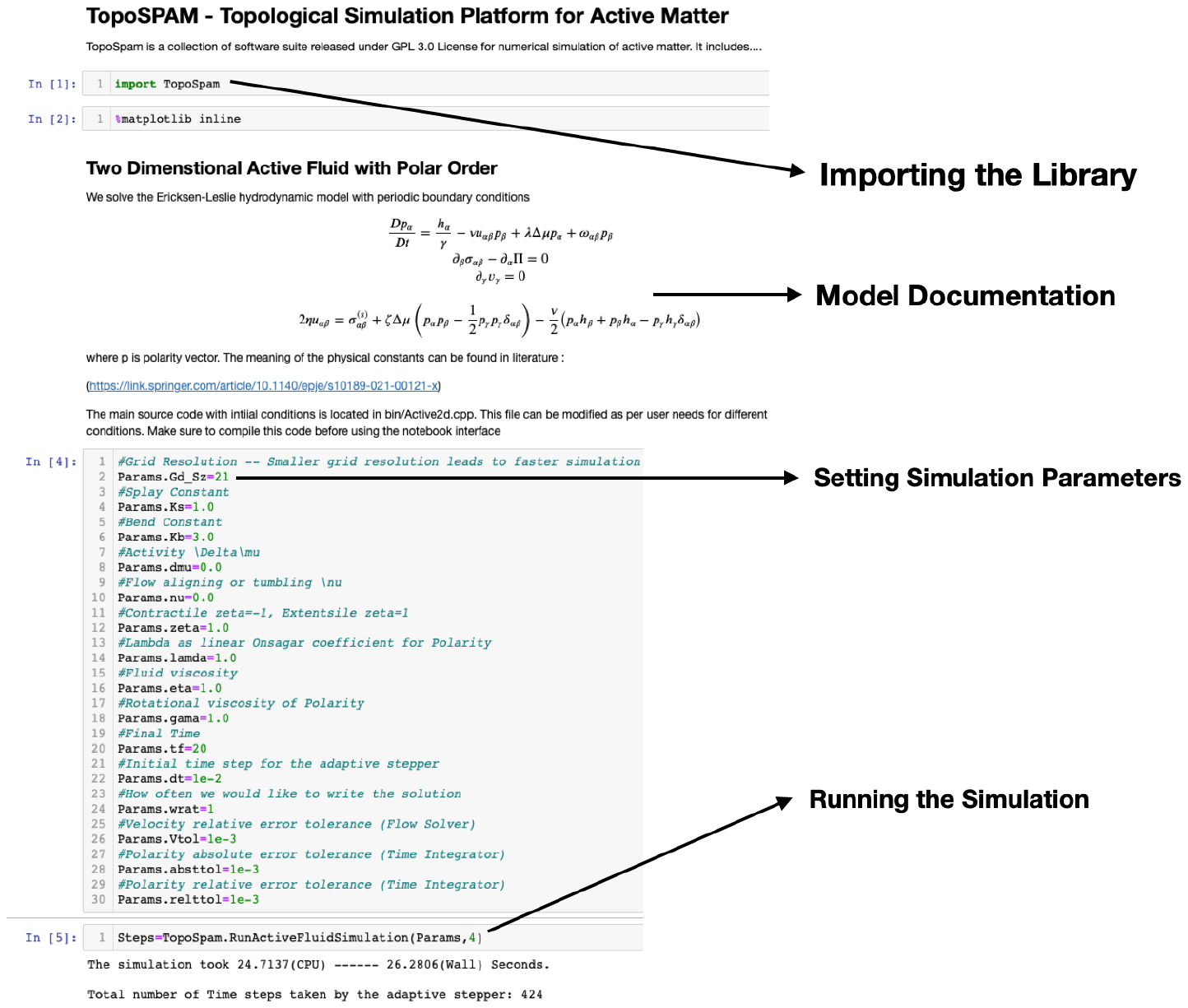}
    \caption{Architecture and user interface of the TopoSPAM software suite. \textbf{(A)} The TopoSPAM architecture is centered around the IPython middleware using Jupyter as the user interface (top). From there, simulations can be set up and run on the scalable OpenFPM high-performance computing platform (left), leveraging its C++ expression system for targeting different hardware backends from laptops to supercomputers (bottom). Simulation results can directly be visualized from within the same Jupyter interface using the state-of-the-art visualization tools ParaView, PyVista, and matplotlib (right). Arrows: The IPython middleware generates the C++ code for OpenFPM, which in turn compiles it to hardware-specific optimized assembly code. Visualization is controlled from the IPython interface, while the data are fetched in the background from the hardware.
    \textbf{(B)} The IPython user interface of TopoSPAM. In this example, the user first imports the TopoSPAM model library and then selects the 2D active fluid model class. The model's documentation is shown inline. Then, the user sets the simulation parameters and runs the model on the target hardware.}
    \label{fig:pipeline}
\end{figure*}

Figure~\ref{fig:pipeline}A illustrates the TopoSPAM architecture, a comprehensive software framework designed for high-performance computing and interactive visualization in computationally studying morphogenesis and active matter. At its core, the architecture leverages high-performance C++ code through the OpenFPM library \cite{Incardona2019} and its custom C++ Template Expression System \cite{Singh2021} for the computationally intensive tasks required by spatio-temporal solvers in 3D+time. OpenFPM and its C++ Template Expression System also ensure portability to different computing hardware as optimized code is automatically generated during compilation from model equations provided in near-mathematical form. This approach differs from existing codes, which are often hand-tuned for a specific model, numerical method, or computer architecture. % and effectively utilizes the Turing-complete nature of the C++ template compiler. 
%there was recent Julia for biologists paper we could cite. but I am not sure if this line belongs here. our focus is more on Spatiotemporal simulations where we leverage openfpm spatial domain decomposition that is not there in other libraries.
Rather, the combination of flexibility and performance is akin to recent developments in Julia \cite{NatureJulia}. The OpenFPM foundation of TopoSPAM, however, does not require manual code tuning, is free of runtime overhead \cite{Incardona:2023a}, and extends to distributed-memory simulations of spatio-temporal models using automatic domain decomposition \cite{Incardona2019}. 
%comes close to such a system for having both flexibility and performance by introducing a dynamic compilation step, but needs hand tuning of the generated assembly by type stabilizing the code. Further, an overhead of compilation or dynamic dispatch is present when running the codes that slows down the computation. 
%CDM: @Ivo, make whatever change, addition, or subtraction here you think is best.
%IFS: done. What do you think?

%Our framework is further supplemented by an independent 3D vertex model solver written in C++ \cite{}.
%IFS: the last sentence should be tied in better. Seems a bit "slammed on".
%AM: that sentence doesn't need to be here at all, I think, as the Vertex model is now also tied into OpenFPM structure (I belive)

This high-performance C++ core interfaces with IPython middleware, as provided by Jupyter, which acts as a flexible and extensible layer for integrating the various components. The IPython middleware can be readily updated to support new codes and functionalities in a modular fashion, enhancing the system's adaptability as shown by the notebook snippet (see Fig.~\ref{fig:pipeline}B). 
%IFS: which notebook snippet? Maybe add figure or equation reference to make clear what is pointed at?

To visualize simulation results, TopoSPAM combines ParaView \cite{Hansen_Johnson_2005} and PyVista 
\cite{sullivan2019pyvista}, coupled with matplotlib \cite{Hunter:2007}. 
%IFS: @Abhinav: please fill in the missing references
%AS: Done
These provide powerful tools for rendering and interactive visualization of scientific data. TopoSPAM seamlessly integrates visualization of results with defining and running simultions under the same IPython interface (see Fig.~\ref{fig:pipeline}B for an example).
%The architecture encompasses both the ``Assembly'' phase, where low-level programming languages create performant simulation codes, and the ``Visualize'' phase for rendering visualization scenes handled modularly. 

Thanks to its performance-portable architecture, TopoSPAM can leverage diverse  hardware platforms, from laptops to high-performance computing clusters. This enables using the same modeling interface to scale from personal devices to large-scale institutional systems, depending on problem requirements. It also allows scientists to seamlessly transition between high-performance simulations and interactive data exploration, providing a versatile platform for complex scientific workflows in fields like biological physics, requiring advanced computation and visualization capabilities.
% (see for example \cite{Singh2023} and the parallel scalability and performance results reported therein).

\subsection{Active polar fluids and the cytoskeleton}

TopoSPAM includes a state-of-the-art method to numerically solve 2D and 3D active polar and nematic hydrodynamic problems in arbitrary geometries and with arbitrary boundary conditions~\cite{Singh2023}. The hydrodynamic equations of active fluids describe the interaction between motile microscopic constituents and their surrounding fluid under energy input. These equations capture many emergent phenomena in biology, for example the dynamics of the cytoskeleton and of active tissue-scale flows~\cite{Marchetti2013}. A mean-field description of such an out-of-equilibrium system has been derived from first principles (conservation laws, symmetries, entropy production, and the Onsager reciprocal relations) and recapitulates the resulting dynamics \cite{kruse2005generic,juelicher2007active,joanny2009active,Juelicher2018}. This model includes two continuous fields: (1) an orientation field to capture the average orientation of the microscopic constituents (e.g., cytoskeletal filaments, motor proteins) at each point in space and time; and (2) a velocity field of the emerging flows and deformations (see Methods section for the governing equations and implementation details).
%IFS: I think these two references are out of place, as they describe active surfaces, whereas all we do in TopoSPAM is bulk.
%(e.g., \cite{Mietke:2018,Mietke:2019})

\begin{figure*}[h]
    \centering
    \includegraphics[angle=-90,width=17cm]{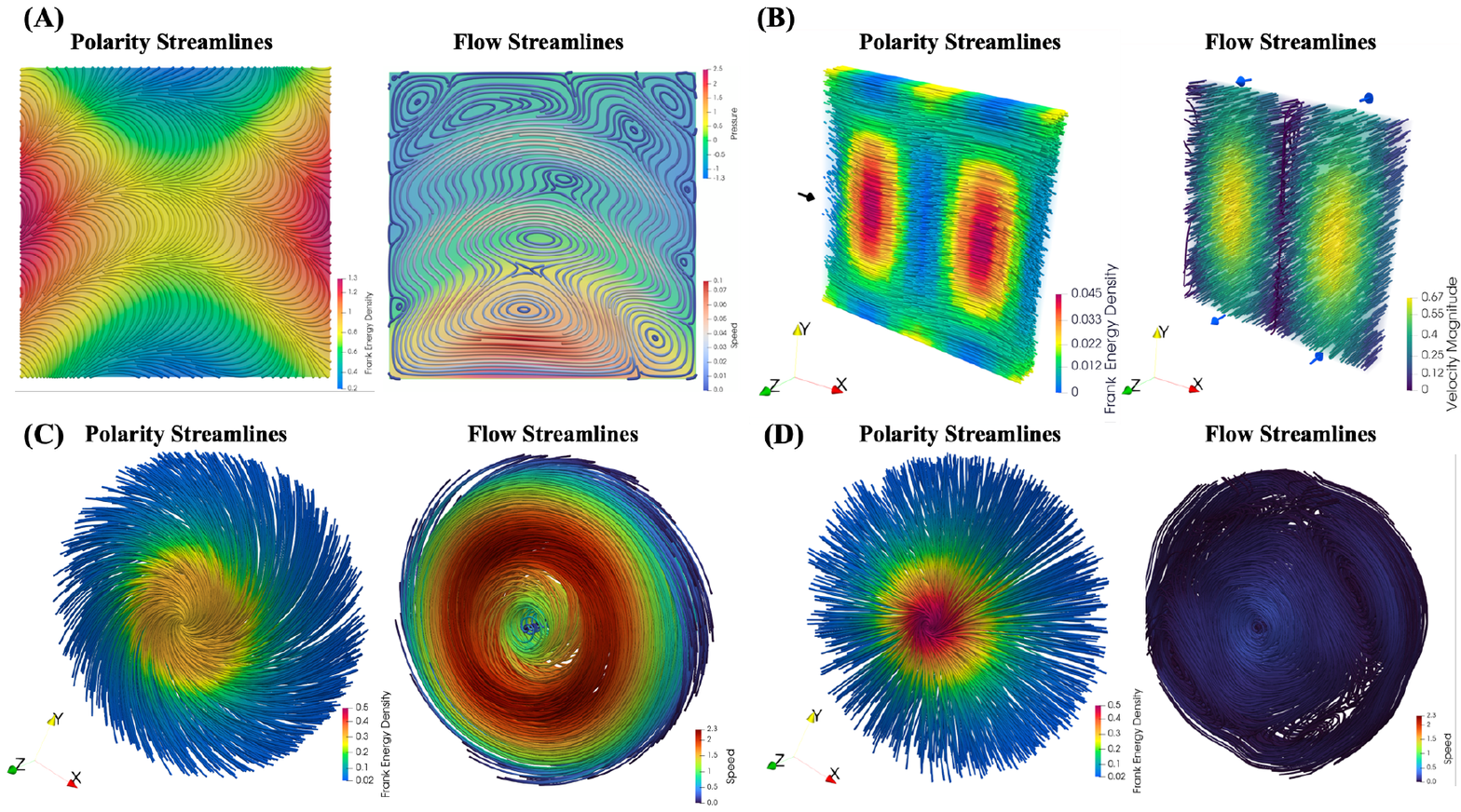}
    \caption{Active polar fluid solutions in 2D and 3D. \textbf{(A)} Two-dimensional active polar fluid with oscillatory polarity field as shown by the polarity streamlines and Frank free energy as color on the left, with resulting flow streamlines and magnitude of the flow velocity as color on the right.
    \textbf{(B)} Three-dimensional extensile active fluid undergoing an out-of-plane bend instability under confinement as shown by the polarity streamlines and Frank free energy as color (left) with flow streamlines and velocity as color (in- and out-of place flow directions indicated by the blue arrows) on the right.
    \textbf{(C)} Simulation of a rotating spiral defect in an active fluid confined to a 3D cylinder.
    \textbf{(D)} Simulation of a stable aster defect with no flow in an active fluid confined to a 3D cylinder.}
    \label{fig:activefluid}
\end{figure*}

Solving such models is numerically challenging but with OpenFPM as its underlying high-performance scientific computing backbone, TopoSPAM streamlines and simplifies this task, achieving efficient scaling on parallel computers (see Ref.~\cite{Singh2023} and the parallel scalability benchmarks therein). A TopoSPAM active fluids simulation is initialized by solving for the steady-state velocity using a given initial polarity field. This initial input may itself be derived from a set of disclination defects in the polarity field. Subsequently, TopoSPAM evolves the polarity field  according to the Lagrangian, co-rotational derivative. This yields the velocity field at the next time point, and TopoSPAM renders the field dynamics by iterating over time steps.

We validated TopoSPAM against the 2D benchmark problem from \cite{Ramaswamy2015}. The results are shown in Fig.~\ref{fig:activefluid}A. This problem represents a test case used to confirm the consistency of numerical methods for active and nematic fluids. The model simulates a thin active polar viscous fluid in a square dish, mimicking an \textit{in vitro} actomyosin film, with no-slip velocity boundary conditions and the filaments anchored at the boundaries. The obtained results are in agreement with previous studies \cite{Ramaswamy2015}.
%IFS: @Abhinav: I think we should include a sentence here saying what the result means. I.e., is it the same as in Ramaswamy et al.? Does it validate? Why/how?
%AS: Added

TopoSPAM's active polar fluid solver has recently been used to confirm the analytically predicted 3D spontaneous flow transition of cytoskeletal suspensions~\cite{AbhinavPRR,Singh2023} and to recapitulate experimentally observed the out-of-plane wrinkling behavior \cite{Alam2024} in microtubule suspensions (Fig.~\ref{fig:activefluid}B).
%IFS: reference missing
%AS: done
Here, we additionally use it to simulate an active polar fluid confined to a 3D cylindrical domain where it develops a moving spiral-shaped line defect under rotational flow (Fig.~\ref{fig:activefluid}C). In the same geometry, a stable aster line defect induces no flow (Fig.~\ref{fig:activefluid}D), highlighting the important role of topology in determining the morphogenetic mode. The last two cases also demonstrate the ability of TopoSPAM to solve complex active fluid problems in non-Cartesian domains.

\subsection{Active vertex models and cells in tissues}

Moving up the scales of biology, TopoSPAM is also capable of capturing discrete cellular behaviors inside an epithelial tissue via its 3D active vertex model framework. A potential approach to linking tissue morphogenesis with mechanistic cell biology involves ``decomposing'' the tissue deformation into distinct, independent cell behaviors, essentially forming its ``deformation basis". Often, these cellular behaviors are intricately interconnected and carefully regulated by numerous mechano-chemical cues to govern the development of tissues and organs, ultimately shaping them into their final forms. Some common cell behaviors include individual cell elongations, T1 rearrangements, cell divisions, apoptosis, cell extrusion, and relative apical--basal surface contractions \cite{JulicherEaton2017,Pearl2017}.

One of the challenges in comprehending and establishing accurate theories of tissue morphogenesis arises from  tissues' ability to modify their internal topology, such as the neighbor network of the cells. This additional layer of complexity makes conventional theoretical frameworks, like linear elasticity and stability analysis, less applicable. To overcome this challenge, a promising approach is to work with models that effectively exploit the discrete topology of the tissue. TopoSPAM provides this through a robust implementations of 3D vertex models and spring-lattice models \cite{vert-models_cont_theory,chaikin1995principles}.

Vertex models provide a powerful framework widely used in tissue mechanics, offering valuable insights into how individual cells' collaborative efforts yield complex responses within a tissue \cite{farhadifar2007influence,staple2010mechanics}. With vertex models, we can understand how cells contribute to shaping the tissue and potentially link deformations to molecular mechanisms.

TopoSPAM also extends this to active vertex models. In contrast to many passive vertex models found in the literature \cite{farhadifar2007influence, vert-models_cont_theory}
our version also considers the active forces that the cells exert on their surroundings. In this way, TopoSPAM's active vertex model enables us to effectively gauge whether active cellular forces or passive relaxation into energetically favorable states drives observed cell and tissue shape changes. 

In particular, TopoSPAM enables taking active forces into account through their effect on cellular traction. For this, TopoSPAM offers multiple coupling choices for connecting traction forces and polarity. In many biologically relevant cases, however, it is convenient to set each cell's traction force to be aligned with the cells' biochemical polarity. Further, traction forces are taken into account by force balance on each vertex to which the standard vertex model work function also contributes. Finally, individual cell's polarity vectors align to each other with a given alignment rate and are subject to rotational noise.

TopoSPAM's active vertex model has recently been used to study experimentally observed cell- and tissue-level behaviors in embryonic mouse pancreas-derived epithelial spheres \cite{Tan:2024}. It had been experimentally observed that such epithelial spheres can perform persistent rotational motion of various types. This tissue-level behavior can be explained by an underlying mechanical interaction between the cells making up the spheroid and the embedding extracellular matrix. These interactions correspond to the cellular traction forces of TopoSPAM's active vertex model. By varying the strength of the traction forces and the polarity alignment rate, TopoSPAM was able to recapitulate all of the experimentally observed modes of tissue movement, including emergent behaviors that were not explicitly input into the model~\cite{Tan:2024}. 

\subsection{Spring lattice models and tissue morphogenesis}

Spring-lattice models take a coarser tissue-level approach where individual cells are no longer resolved. They focus on mean or collective, instead of individual, cellular behavior. The versatility of spring lattice models allows them to capture various morphogenetic events, such as tissue folding or invaginations~\cite{Vellutini2023.03.30.534554}. In addition, TopoSPAM's spring lattice model can effectively capture convergent-extension deformations (compare \cite{WALLINGFORD2002695}) and  metric shape programming by coarse-graining over patterned T1 events \cite{FuhrmannKrishna2024}.

To effectively use the spring lattice model in TopoSPAM, a sensible mapping between the average cell behavior and the spontaneous strains in the tissue must be specified. This can be achieved through phenomenological models describing the process of interest or by solving the required shape inverse problem while considering biological constraints.

\begin{figure*}[h]
    \centering
    \includegraphics[width=17cm]{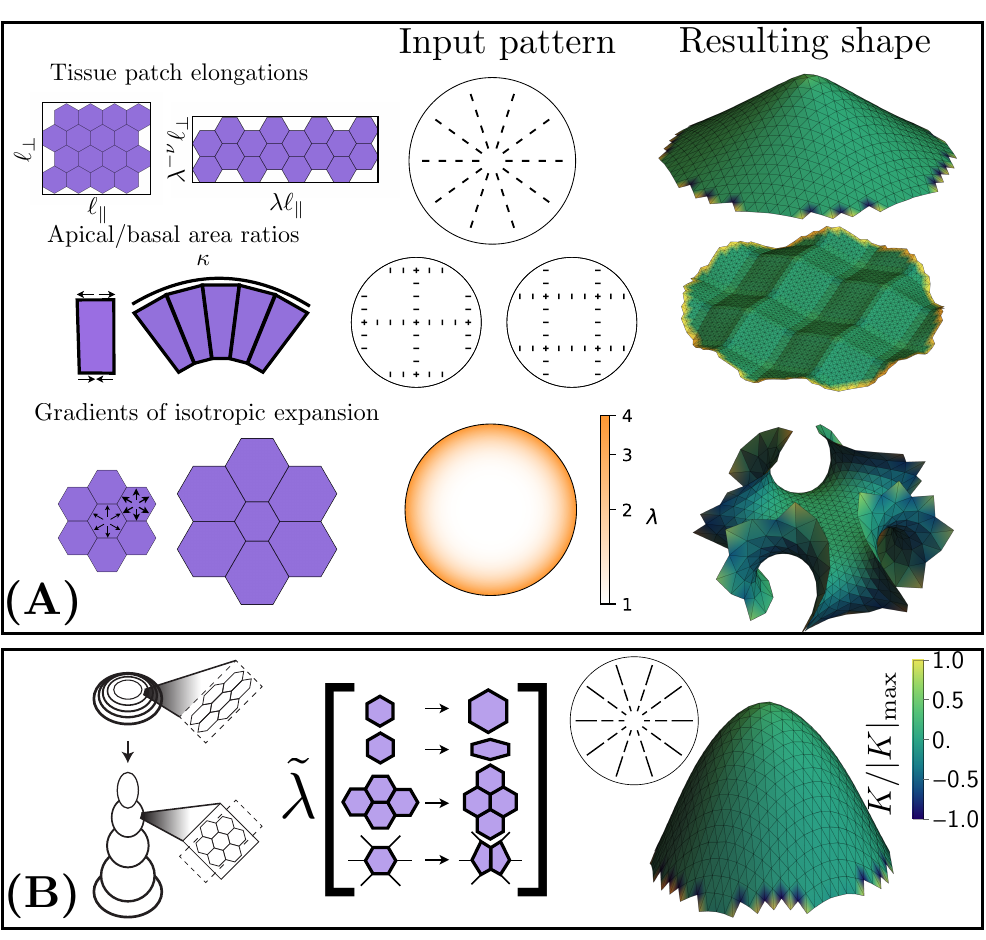}
    \caption{\textbf{(A)} Overview of the spring-lattice models for morphogenesis: Left column: Shape response of a group of cells to a specific collective cellular behavior. Middle column: coarse-grained shape patterns using the shape mechanisms of the left as building blocks. Right column: 3D shape obtained after letting the spring-lattice system relax to a geometrical configuration that satisfies force balance for the rest lengths imposed by their corresponding planar deformation pattern. From top to bottom: uniform radial elongations yield conical surfaces. Ratios of apical and basal areas manipulated to program a corrugated surface. Axisymmetric gradients of cell expansion decreasing along the radial direction yield a hyperbolic surface. \textbf{(B)} The shape development of the {\it Drosophila} imaginal leg disc elongation is an example of developmental morphogenesis (left) that could be captured by a spontaneous-strain deformation tensor that combines different ``deformation modes'' (middle). Radial elongation gradients transform a flat disk into an elongated dome. Each of the shown 3D surfaces is colored with respect to their Gaussian curvature (color bar). %\cmd{permissions to add the figure on the bottom left might be needed}
    }
    %IFS: please check if permission is needed.
    \label{fig:vm-sl}
\end{figure*}

In Fig.~\ref{fig:vm-sl}A, we illustrate some shape-encoding building blocks that can be used within the TopoSPAM spring lattice model. For example, multiple T1 rearrangements along a preferred axis, together with the assumption that the tissue remains confluent, lead to tissue patches of anisotropic, elongated shape. Patterns of such elongations can create geometric incompatibility in the plane of the tissue, which in turn leads to out-of-plane deformation and the establishment of new 3D ``ground state" shapes. The details of the deformations along the principal directions of the tissue patch generally depend on the specific cellular processes considered.

A spring lattice model does not consider how the internal cell-neighborhood topology of the patch is reorganized, but it instead focuses on the overall behavior of the patch as a whole. An equivalent mean response can, for example, be achieved through cell divisions by specifying a nearly-fixed division axis, or by keeping the patch network fixed and elastically deforming each cell in the patch. The collective effect in both cases is the same. 

Thanks to their simplicity, spring lattice models can be extended to situations where the apical and basal surfaces of the tissue differ or have distinct preferred elongation axes. An example of this strategy's success is the encoding of a preferred tissue curvature. The center column of Fig.~\ref{fig:vm-sl}A shows different deformation modes projected onto the spring rest lengths of the model. Although circular tissue geometries are shown for illustration purposes, TopoSPAM allows these to be modified to more general cases.
In the top row of Fig.~\ref{fig:vm-sl}A, for example, a radial elongation pattern results in the formation of a conic surface~\cite{PhysRevE.84.021711}. In the middle row, area ratio and preferred elongation are independently imposed on subsets of cells, leading to a corrugated surface~\cite{corrugated_surface}. Finally, in the bottom row, a radial swelling gradient creates the complex hyperbolic surface on the right \cite{Kim-science-2012}. Even though in a living tissue, the source of deformation is not usually described as ``programmed'', it can still be interpreted as such.

TopoSPAM's spring lattice model has recently been used to understand the complex shaping of the {\it Drosophila} fly wing during the morphogenetic process of eversion \cite{FuhrmannKrishna2024}, and in Fig.~\ref{fig:vm-sl}B we provide a similar example of how the TopoSPAM model could capture the shape development of {\it Drosophila} imaginal leg disc. In this precursor tissue of legs in {\it Drosophila}, it has been observed that cells are elongated along azimuthal directions~\cite{legdisk}. It has been proposed that upon a hormonal cue, these cells take up isotropic shapes, which would lead to radial elongation and azimuthal contraction, allowing a flat tissue to evaginate into a cone or tube-like structure. Thus, the cellular packing in this tissue has programmed a specific deformation pattern that can be switched on at the correct developmental time. Similarly, it has been shown that there is a spatial pattern of apical cell behaviors in the wing precursors (wing imaginal discs) of {\it Drosophila}~\cite{10.7554/eLife.57964}.

Ultimately, the versatility of the spring lattice model allows one to specify any spontaneous deformation gradient tensor that is a combination of modes of the deformation basis of morphogenesis.

%In this case uniaxial elongations gradients along the radial direction of the initially flat disk provide a minimal model that is sufficient to capture the main development features of leg disc elongation. \cmd{AK: can you confirm if that this deformation pattern accurately captures the leg-disk development?}.

\section*{Discussion}

We presented an integrated, topology grounded simulation platform for morphogenesis and biological active matter, named TopoSPAM~\cite{TopoSPAM}. The presented software platform combines high-performance meshfree particle methods for 2D and 3D continuum models, with a quasi-3D active vertex model and spring lattice models for coarse-grained dynamics into a coherent multiscale simulation framework under a common Python user interface. This enables multiscale simulations and modeling of morphogenetic processes and active biology from molecules to cells to tissues with integrated results visualization.

We have shown how TopoSPAM can be used to numerically solve a range of problems ranging from cytoskeletal hydrodynamics to active tissue mechanics. For active hydrodynamics, this specifically included 2D and 3D cytoskeletal-cortex like channel flows, where we have simulated the spontaneous onset of flows above a certain chemical activity and an out-of-plane wrinkling instability. For tissue mechanics, this included complex couplings between active cellular traction forces and emergent tissue behaviors, as well as coarse-grained acquisition of complex 3D shapes during strain-mediated tissue morphogenesis. All of these examples have been experimentally observed in the literature and theoretically validated. 

With its combined capabilities, TopoSPAM represents a powerful methodological advance. But it is not without its limitations. First, understanding morphogenesis and biological active matter is a fundamentally multiscale problem. While TopoSPAM includes methods and models at various mesoscales, it is not yet capable of performing fully coupled multiscale modeling in the sense that simulations performed at these different scales and with the different models cannot be nested. Future work could implement time-splitting techniques \cite{Samaey2005-ck}
%IFS: @Abhinav: include missing reference.
%AS: Done
or computational coarse-graining \cite{Supekar2023}
%IFS: include missing reference
to couple dynamics across scales. Second, the underlying physics and biology are still partly unknown for many morphogenetic processes. Out of the box, TopoSPAM provides solvers for the known governing equations. As our understanding of morphogenesis grows, one may need to change these equations or otherwise supplement the simulated physics, which requires editing and recompiling the source code. Third, many biological processes are inherently stochastic, causing phenotypic variability during morphogenesis. While the models in TopoSPAM can account for extrinsic, imposed noise, they cannot, however, represent intrinsic, coupled noise (as in stochastic differential equations). Future work could allow for stochasticity within each of the core TopoSPAM models, further expanding its capabilities and applicability.

Despite these potential future developments, TopoSPAM already is a powerful and useful tool with significant potential to aid our understanding of mesoscale biological physics. On top of its many features, TopoSPAM has the added advantage that it is an integrated platform with a single, coherent user interface, improving ease of use and accessibility for difficult multi-context problems. By folding topological considerations into many of the current cutting-edge computational models and packaging them together, TopoSPAM is poised to be an invaluable asset across biological physics and biology as a whole.

\section*{Methods}
\setcounter{subsection}{0}

TopoSPAM is a suite of custom software on the basis of the high-performance scalable scientific computing library OpenFPM \cite{Incardona2019}. The structure of TopoSPAM is visualized in Fig.~\ref{fig:pipeline}A.
We use the Jupyter Ipython environment as a middleware to run the C++ high-performace computer codes on parallel and distributed computers. The simulations are visualized on the same computers using the TopoSPAM Python interface for ParaView and PyVista. In Fig.~\ref{fig:pipeline}B, we show an example of the user interface. In this example, the user imports the TopoSPAM model library and then selects one of the available 2D active fluid models. TopoSPAM provides inline documentation and explanation of the selected model class. The user then sets the model parameters and runs and visualizes the simulation with a single function call. This function call internally launches the C++ code of a high-performance simulation without Python being a bottleneck for intensive numerical computations.
Given the open-source nature and underlying C++ expression system \cite{Singh2021}, it is straightforward to modify the model library for custom needs and new emerging models. TopoSPAM's modular, polylithic architecture allows users to write custom models and seamlessly add them, and their documentation, to the TopoSPAM IPython interface.

%IFS: This figure was not referenced or used anywhere in the text, so I commented it out. Feel free to bring it back and add the required text discussing it in the appropriate place.
%\begin{figure}[h]
%    \centering
%    \includegraphics[width=13.5cm]{figs/SpringLattice_notebook.png}
%    \caption{Screenshot for a notebook showing how to run a spring lattice simulation}
%    \label{fig:notebook-sl}
%\end{figure}

Currently, TopoSPAM ships with the built-in models described below. All models are available in 2D and 3D, and many of them as both passive or active variants.

\subsection{Polar Active Hydrodynamics}

TopoSPAM includes the 2D and 3D active and passive Ericksen-Leslie equations modeling an incompressible polar fluid in the hydrodynamic limit~\cite{Juelicher2018,Marchetti2013}. This model describes the behavior of viscous active polar fluids under non-equilibrium, nonlinear hydrodynamic conditions. It relates the dynamics of two fields: (1) the macroscopic orientation field $\mathbf{p}$ with components $p_{\alpha}$ defined by the average orientation of the microscopic constituents in an infinitesimal volume; (2) the velocity field $\mathbf{v}$ with components $v_{\alpha}$ of the emerging flows and deformations. The governing Stokes equation for the force balance is
\begin{equation}
	\partial_{\beta} \sigma_{\alpha \beta}^{(\text{tot})}-\partial_{\alpha} \Pi=0\, ,
	\label{eq::activeGel_Stokes}
\end{equation}
with $\alpha,\beta,\gamma \in \{x,y,z\}$ the spatial vector components, $\sigma_{\alpha\beta}^{(\text{tot})}$ the total deviatoric stress tensor and $\Pi$ the pressure. Repeated indices imply summation (Einstein summation notation). Incompressibility of the solvent is imposed by $\partial_{\gamma}v_{\gamma} = 0$. We decompose the total deviatoric stress tensor into its symmetric part $\sigma_{\alpha \beta}^{(s)}$, its antisymmetric part $\sigma_{\alpha \beta}^{(a)}$, and the Ericksen stress tensor $\sigma_{\alpha \beta}^{(e)}$. The symmetric stresses $\sigma_{\alpha \beta}^{(s)}$ are defined by the constitutive relation 
\begin{align}
	\sigma_{\alpha \beta}^{(s)} =2 \eta u_{\alpha \beta} +\zeta \Delta \mu\left(p_{\alpha} p_{\beta}-\frac{1}{3} p_{\gamma} p_{\gamma} \delta_{\alpha \beta}\right) \nonumber \\
	+\frac{\nu}{2}\left(p_{\alpha} h_{\beta}+p_{\beta} h_{\alpha}-\frac{2}{3} p_{\gamma} h_{\gamma} \delta_{\alpha \beta}\right)\, , 
	\label{eq::activeGel_constitutive}
\end{align}
whereas $\sigma_{\alpha \beta}^{(a)}$ and $\sigma_{\alpha \beta}^{(e)}$ are given by
\begin{equation}
	\sigma_{\alpha\beta}^{(a)} = \frac{1}{2} (p_{\alpha} h_{\beta} - p_{\beta} h_{\alpha}) \, ,\qquad 	\sigma_{\alpha\beta}^{(e)} = - \frac{\partial f}{\partial (\partial_{\beta}p_{\gamma})} \partial_{\alpha}p_{\gamma} \, .
	\label{eq:activeGel_antisymStress}
\end{equation}

Equation~\ref{eq::activeGel_constitutive} relates the strain-rate tensor $u_{\alpha \beta}=\frac{1}{2}\left(\partial_{\alpha} v_{\beta}+\partial_{\beta} v_{\alpha}\right)$ to the symmetric, the active, and the passive stresses. The material parameter $\eta$ is the fluid viscosity, $\zeta$ is the coupling coefficient linking the material stress to the chemical potential difference $\Delta\mu$ ($\zeta<0$ contractile stress, $\zeta>0$ extensile stress), $\nu$ is a coupling coefficient linking the mechanical stress to the orientation field ($|\nu|>1$ flow-aligning, $|\nu|<1$ flow-tumbling), and $\mathbf{h}$ is the so-called {\em molecular field} which is the conjugate force to the orientation field $\mathbf{p}$ and defined as the functional derivative $h_{\alpha}=-\delta F / \delta p_{\alpha}=-\partial f / \partial p_{\alpha} + \partial_{\beta}(\partial f / \partial(\partial_{\beta}p_{\alpha}))$ of the {\em Frank free energy} $F$, which in 3D is:
\begin{align}
	F_{3D}=\int \Big[ &\frac{K_s}{2} (\partial_\alpha p_\alpha)^2+ \frac{K_t}{2} (p_\alpha \varepsilon_{\alpha\beta\gamma} \partial_\beta p_\gamma)^2 +  \nonumber\\
	&\frac{K_b}{2}  ( \varepsilon_{\delta\varepsilon\alpha} p_\varepsilon (\varepsilon_{\alpha\beta\gamma} \partial_\beta p_\gamma))^2 - \frac{1}{2}h^0_{\Vert} p_\alpha p_\alpha \Big]\,  \mathrm{d}V\, ,
	\label{eq:freeEnergy}
\end{align}
where $K_s$, $K_t$, and $K_b$ denote the splay, twist, and bend elasticity constants of the material, respectively, and $\varepsilon_{\alpha\beta\gamma}$ is the Levi-Civita symbol. The whole integrand is referred to as the {\em Frank free-energy density} $f$. The term $h^0_{\parallel}$ serves as Lagrange multiplier to constrain the orientation field to unit magnitude $\|\mathbf{p}\| = 1$, which allows for the presence and emergence of topological defects. It is derived from the co-rotational Lagrangian derivative 
\begin{equation}
	\frac{\mathrm{D} p_{\alpha}}{\mathrm{D} t} = \frac{\partial p_{\alpha}}{\partial t}+v_{\gamma} \partial_{\gamma} p_{\alpha}+\omega_{\alpha \beta} p_{\beta} = \frac{h_{\alpha}}{\gamma}-\nu u_{\alpha \beta} p_{\beta}+\lambda\Delta\mu p_\alpha
	\label{eq:activeGel_LagrangianDerivative}
\end{equation}
by enforcing  $p_\gamma\frac{Dp_\gamma}{Dt}=0$. The parameter $\gamma$ denotes the rotational viscosity of the polarity field and $\omega_{\alpha \beta}=\frac{1}{2}\left(\partial_{\alpha} v_{\beta}-\partial_{\beta} v_{\alpha}\right)$ is the vorticity tensor. Taking the divergence of $\sigma_{\alpha \beta}^{(\text{tot})}$ leads to the final nonlinear Stokes equation.

TopoSPAM numerically solves these nonlinear equations using a fully verified and convergence-validated numerical method~\cite{Singh2023}. The solver handles both 2D and 3D simulation domains of arbitrary shape and with arbitrary (Neumann, Dirichlet, Robin) boundary conditions for the polarity and velocity fields. 
For the numerical solution, the continuous fields are discretized on computational particles. This avoids having to generate a grid or mesh and better extends to complex and moving geometries, as prevalent in morphogenesis. Differential operators are consistently approximated over the resulting (dynamic) collocation points using the meshless DC-PSE method~\cite{Schrader2010}. The implicit equation for the velocity is directly solved on the irregularly distributed particles using the KSPGMRES linear system solver from PETSc~\cite{balay_efficient_1997}.

The time dynamics of the fields is resolved using the adaptive Adams--Bashforth--Moulten predictor-corrector time stepping method of convergence order two. The entire active fluid simulation code is written in the OpenFPM C++ expression system \cite{Singh2021} in only 423 lines of C++ source code that transparently parallelizes to shared- and distributed-memory computers and to Graphics Processing Units (GPUs) \cite{Singh2023,Incardona:2023a}.

\subsection{Active Vertex Models}

As described in more detail elsewhere \cite{Tan:2024}, the quasi-3D vertex model included in the TopoSPAM package is an active vertex model on a spherical surface. As such, it represents individual cells as polygons and makes use of the standard vertex model work function
\begin{equation}
    W = \sum_{\alpha\in \text{cells}} \frac{1}{2}K^\alpha \left( A^\alpha - A_0^\alpha \right)^2 + \sum_{\alpha\in \text{cells}} \frac{1}{2}\Lambda^\alpha L^\alpha ,
\end{equation}
where $K^\alpha$ is the area stiffness, $A_0^\alpha$ the preferred cell area, $\Lambda^\alpha$ the perimeter tension magnitude, and $L^\alpha$ the perimeter of cell $\alpha$. 

In addition, cells in our vertex model are equipped with a polarity vector $\mathbf{p}_\alpha$ for each cell $\alpha$. The dynamics of $\mathbf{p}_\alpha$ is given by its co-rotational time derivative $D/Dt$ as
\begin{equation}
    \frac{D \mathbf{p}_\alpha}{D t} = \gamma\left< \mathbf{p} \right>_\alpha + \sqrt{2 \mathcal{D}_r} {\bm \eta}(t) + \mu(t) \mathbf{p}_\alpha + p_\alpha^n \hat{\mathbf{n}}_\alpha \, .
    \label{eq:vertexmodel_polaritydynamics}
\end{equation}
Here, $\gamma$ is the alignment rate of the polarity of cell $\alpha$ with the average polarity of its nearest neighbors
\begin{equation}
    \left< \mathbf{p} \right>_\alpha = \frac{1}{M_\alpha} \sum_{\left< \alpha' | \alpha \right>} \mathbf{p}_{\alpha'}\, ,
\end{equation}
with $M_\alpha$ being the number of vertices of cell $\alpha$. Further, in Eq.~\ref{eq:vertexmodel_polaritydynamics}, $\mathcal{D}_r$ is a rotational diffusion constant and ${\bm \eta} (t)$ a white noise 
%IFS: in the Conclusions, we say that one limitaiton of TopoSPAM is that is cannot model stochastic effects. But then there is white noise here. Seems a contradiction? Maybe the Conclusions need to be adjusted?
oriented perpendicular to the current cell polarity in the tangent plane of the sphere, i.e. ${\bm \eta}(t) = \hat{\mathbf{s}}_{\perp} \eta(t)$ with $\hat{\mathbf{s}}_{\perp} = \hat{\mathbf{n}}_\alpha \times \mathbf{p}_\alpha / |\hat{\mathbf{n}}_\alpha \times \mathbf{p}_\alpha|$ and $\hat{\mathbf{n}}_\alpha$ being the outward normal. Finally, Eq.~\ref{eq:vertexmodel_polaritydynamics} also contains terms to impose $|\mathbf{p}_\alpha| =1$, as in the continuum active hydrodynamics model, through the Lagrange multiplier $\mu(t) = - \gamma \mathbf{p}_\alpha \cdot \left< \mathbf{p} \right>_\alpha$ and to keep $\mathbf{p}_\alpha$ in the tangent plane of the sphere by means of adding the normal $p_\alpha^n \hat{\mathbf{n}}_\alpha$ with $p_\alpha^n = - \gamma \left< \mathbf{p} \right>_\alpha \cdot \hat{\mathbf{n}}_\alpha$.

The thus defined polarity field $\mathbf{p}$ underlies the active cellular forces that distinguish the TopoSPAM vertex model from many other vertex models. We propose that any cellular traction force $\mathbf{f}_{m}^{\text{active}}$ acting on a vertex $m$ in the model acts in a direction determined by the polarities of the cells abutting in vertex $m$ as
\begin{equation}
    \mathbf{f}_m^{\text{active}} = F \sum_{\left< \alpha | m \right>} \frac{\mathbf{p}_\alpha}{M_\alpha} \, ,
\end{equation}
with traction force magnitude $F$ and normal component 
\begin{equation}
f_m^n + \left( \frac{\partial W}{\partial \mathbf{X}_m} - \mathbf{f}_m^{\text{active}} \right) \cdot \hat{\mathbf{n}}_m \, .
\end{equation}
Here, $\mathbf{X}_m$ is the position of vertex $m$ and $\hat{\mathbf{n}}_m$ the outward normal. 

Combining this with the standard vertex model work function, the active cellular forces and the friction given by the velocity of vertex m, $\mathbf{v}_m$, and a friction coefficient $\xi$ with the surrounding material yields the force balance equation of TopoSPAM's active vertex model
\begin{equation}
    \xi \mathbf{v}_m = \mathbf{f}_m^{\text{active}} - \frac{\partial W}{ \partial \mathbf{X}_m } + f_m^n \hat{\mathbf{n}}_m \, ,
\end{equation}
according to which the system evolves in time.

\subsection{Spring Lattice Models}

This class of TopoSPAM models describes elastic thin (2D) or thick (3D) tissues. The simulation lattice consists of vertices on the top and bottom surfaces of these. Vertices can be spatially distributed in the two surfaces either as a regular lattice or randomly using algorithms like Poisson disk sampling~\cite{bridson2007fast}. Vertices are connected to nearby vertices using different algorithms like Delauney triangulation. In a 3D model, vertices are connected within the plane as well as through the thickness of the sheet. In a 2D model, the elastic sheet can be either a plane or a curved manifold. 

The edges connecting neighboring vertices act as elastic springs with either extrinsic or intrinsic viscosity. When a lattice is initialized in this way, each spring is assigned to be stress-free: the initial rest length of each spring at time $t = 0$, $\delta^o_i$, is the distance between the two vertices that it connects, 
\begin{equation}
    \delta_{i}^{o}= |\Delta \mathbf{X}| \, , 
\end{equation}
where $\Delta \mathbf{X}$ is the difference between the position vectors of the two endpoints. 

To implement shape change, we use a spontaneous deformation tensor $\boldsymbol{\lambda}$, which acts on the springs to compute the new rest length of each spring, $\delta^o_f$, as
\begin{equation}
    \delta^o_f = |\boldsymbol{\lambda}\cdot \Delta \mathbf{X}| \, .
\end{equation}
Assigning the new rest lengths to the springs in the lattice leads to stresses, which relax by displacing the vertices until force balance is achieved. This is computed using the classic gradient descent algorithm.

This model allows for a continuous field of deformation tensors, $\boldsymbol{\lambda}(\mathbf{x})$, which is evaluated at the two endpoints of each spring. The deformation field, $\boldsymbol{\lambda}(\mathbf{x})$, is a $3 \times 3$ tensor in which each component is the multiplicative factor by which the rest length is changed in that direction. 

Such deformation tensor fields are used extensively in the literature of shape programmable materials \cite{modes2016,warner2007liquid}. Sheets of exotic materials, like liquid crystal elastomers, can be designed such that, upon being exposed to a cue, they deform in different directions by different amounts~\cite{herbert2022synthesis}. This pattern of programmed deformation is  captured by a spontaneous deformation field. Such a tensor has also been used previously to describe growth patterns in living tissues~\cite{BENAMAR20052284}. 

In TopoSPAM, the user specifies the deformation field $\boldsymbol{\lambda}(\mathbf{x})$ component wise. The model denotes the direction perpendicular to the nematic director $\mathbf{p}$ and tangent to the surface by $\mathbf{n}^*$ and the direction along the height/thickness by $\mathbf{t}$. All vectors $\mathbf{p}, \mathbf{n}^*$, and $\mathbf{t}$ are unit vectors. To deform the surface along these directions by multiplicative factors of $\lambda_p$, $\lambda_{n^*}$, and $\lambda_t$, respectively (which can all be functions of space), the deformation field $\boldsymbol{\lambda}(\mathbf{x})$ is given by
\begin{equation}
    \boldsymbol{\lambda}(\mathbf{x}) = \lambda_p (\mathbf{p} \otimes \mathbf{p}) + \lambda_{n^*} (\mathbf{n}^* \otimes \mathbf{n}^*) + \lambda_t (\mathbf{t} \otimes \mathbf{t})\, .
    \label{eq:springlatticemodel_deformationfield}
\end{equation}
Applying user specified $\mathbf{p}$, $\lambda_p$, $\lambda_{n^*}$, and $\lambda_t$ as function of space (with $\mathbf{p}$ tangent to the surface of the lattice), the code automatically computes the spontaneous deformation field using equation \ref{eq:springlatticemodel_deformationfield}.

As an example, isotropic swelling in the surface by a factor of $\lambda$ and an anisotropic deformation along $\mathbf{p}$ by a factor of $\tilde{\lambda}$ with a spontaneous Poisson factor of $\nu$ would be specified as:
\begin{equation}
    \begin{split}
        \lambda_p &= \lambda \tilde{\lambda} ,\\
        \lambda_{n^*} &= \lambda \tilde{\lambda}^{-\nu} ,\\
        \lambda_t &= 1   \, . 
    \end{split}
\end{equation}

\section*{Acknowledgments}
The authors acknowledge support by the following funding bodies: Federal Ministry of Education and Research of Germany (Bundesministerium f\"{u}r Bildung und Forschung, BMBF) funding codes: 031L0160 (project ``SPlaT-DM -- computer simulation platform for topology-driven morphogenesis'') (authors A.Si., A.A., P.I., C.M.D., M.B.), Center for Scalable Data Analytics and Artificial Intelligence (ScaDS.AI) Dresden/Leipzig (author Ab.S.), 01EJ2206A (project ``PACETherapy – Pathological cell extrusion therapy'' (author P.H.S.); Center for Advanced Systems Understanding (CASUS), financed by Germany’s Federal Ministry of Education and Research (BMBF) and by the Saxon Ministry for Science, Culture and Tourism (SMWK) with tax funds on the basis of the budget approved by the Saxon State Parliament (author S.K.T.V.); and the European Union’s Horizon 2020 Research and Innovation Programme under grant agreement no. 829010 (PRIME) (authors A.Sz., M.B.).

\bibliographystyle{unsrt}
\bibliography{references.bib}

\end{document}